\documentclass[preprint2]{aastex}

\begin{document}

\title{A search for Extraterrestrial Intelligence (SETI) toward the Galactic Anticenter with the Murchison Widefield Array}

\author{S.J. Tingay\altaffilmark{1}, C.D. Tremblay\altaffilmark{1}, \& S. Croft\altaffilmark{2}}

\altaffiltext{1}{International Centre for Radio Astronomy Research (ICRAR), Curtin University, Bentley, WA 6102, Australia}
\altaffiltext{2}{Department of Astronomy, University of California Berkeley, Berkeley, CA, USA}


\begin{abstract}
Following from the results of the first systematic modern low frequency Search for Extraterrestrial Intelligence (SETI) using the Murchison Widefield Array (MWA), which was directed toward a Galactic Center field, we report a second survey toward a Galactic Anticenter field.  Using the MWA in the frequency range of 99 to 122 MHz over a three hour period, a 625 sq. deg. field centered on Orion KL (in the general direction of the Galactic Anticenter) was observed with a frequency resolution of 10 kHz.  Within this field, 22 exoplanets are known.  At the positions of these exoplanets, we searched for narrow band signals consistent with radio transmissions from intelligent civilisations.  No such signals were found with a 5$\sigma$ detection threshold.  Our sample is significantly different to the 45 exoplanets previously studied with the MWA toward the Galactic Center \citep{tin16}, since the Galactic Center sample is dominated by exoplanets detected using microlensing, hence at much larger distances compared to the exoplants toward the Anticenter, found via radial velocity and transit detection methods.  Our average effective sensitivity to extraterrestrial transmiter power is therefore much improved for the Anticenter sample.  Added to this, our data processing techniques have improved, reducing our observational errors, leading to our best detection limit being reduced by approximately a factor of four compared to our previously published results.
\end{abstract} 


\section{INTRODUCTION}
This paper continues efforts to utilise the Murchison Widefield Array (MWA: \citealt{tin13}) for the Search for Extraterrestrial Intelligence (SETI).  Previously, we utilised MWA observations primarily targetted at the detection of astrophysical spectral lines along the Galactic Plane, toward the Galactic Centre, over a 400 sq. deg. field, to undertake an opportunistic search for narrow band SETI signals in the frequency range 103 to 133 MHz \citep{tin16}.  The reader can refer to \citet{tin16} for the general context of SETI research at radio wavelengths, as well as some brief history specifically for low frequency radio observations.  We will not repeat this context here.  An up-to-date description of SETI surveys conducted at cm-wavelengths can be found in \citet{enr17} and an excellent graphical summary of previous radio SETI experiments can be found in \citet{gra17} (their Figure 9).

To briefly reiterate the unique capabilities of the MWA described in \citet{tin16}, we cover a frequency range for SETI that has been poorly explored to date (in particular from a pristine radio quiet environment) and the MWA has an extremely wide field of view (up to 1000 sq. deg., depending on frequency).  This means that very large numbers of SETI targets can be examined in any given observation in a unique frequency range.  \citet{gra17} place the \citet{tin16} results in the overall context of SETI experiments and show that they are highly competitive in this frequency range.

We did not detect any SETI signals from the 45 exoplanets known in our previous Galactic Centre field, with a best limit on estimated Equivalent Isotropic Radiated Power (EIRP) of approximately $4\times10^{13}$ W from GJ 6676c at a distance of 6.8 pc.  Only 4/45 exoplanets in our Galactic Centre sample are closer than 50 pc.  The remaining 41 exoplanets are all more distant than 1 kpc, since the sample is dominated by exoplanets detected using microlensing techniques.  Therefore, the vast majority of EIRP limits in the Galactic Center field sample are more than four orders of magnitude higher than our best limit in that field.

In this paper, we examine a different field with the MWA, generally toward the Galactic Anticenter (centred on Orion KL), where the known exoplanets are mainly detected using radial velocity or transit techniques.  The 22 known exoplanets in this field consist of 15/22 closer than 100 pc with 12/22 closer than 50 pc.  The closest exoplanets in this field are BD-06 1339b/c, at a distance of 5.32 pc, 20\% closer than the closest exoplanet in our Galactic Center field.  Thus, on average, our sensitivity to extraterrestrial transmitted radio power is much higher for our observations in the Galactic Anticenter field, compared to the Galactic Center field.

The MWA is participating in a new wave of SETI experiments of various types, summarised (MWA included) in \citet{wor17} and largely driven by the new Breakthrough Listen program.  Breakthrough Listen is gaining momemtum and has published the first results from a survey of 692 nearby stars between 1.1 and 1.9 GHz with the Green Bank Telescope (GBT) \citep{enr17}.  The upper limits on EIRP from these early BL results are of the order $10^{13}$ W, comparable to the limits we present in \citet{tin16} from the MWA.  Thus, interestingly, the best GBT and MWA limits are very similar but cover frequencies that are an order of magnitude different (100 - 200 MHz vs 1 - 2 GHz).  There is no overlap between the 692 targets of \citet{enr17} and the sample of 22 exoplanets reported here.

Also, the GBT survey searched for signals with a bandwidth of a few Hz, whereas our survey has a resolution of 10\,kHz. The signal-to-noise ratio of a narrow-band signal (if corrected for Doppler drift) will be much lower in a 10\,kHz band than it will be in one with resolution that approximately matches the bandwidth of the signal. However, although transmitter power requirements might influence an extraterrestrial civilization to prefer to broadcast narrow-band signals, it is by no means certain that they would choose to do so. For example, spread spectrum (wide band) communications techniques have been considered in the past by some authors \citep{mes12}.  In addition, the coarser frequency resolution of our survey means that a typical signal will stay confined to a single channel during the integration time of our observations, meaning that we can simply look for outliers in individual channels, rather than performing a full Doppler drift search. Experiments with the MWA are therefore complementary to those with the GBT or other facilities at much higher frequency resolutions.

In the following sections we describe our observations and data analysis, and provide discussion of our conclusions.

\section{OBSERVATIONS AND DATA ANALYSIS}

Observations with the MWA took place on 21 November 2015, centered on the Orion Kleinmann-Low Nebula (Orion KL), as described in Table \ref{obs}. Dual-polarisation data were obtained in a 30.72 MHz contiguous bandwidth.  A two stage polyphase filterbank splits the bandwidth into 24$\times$ 1.28\,MHz ``coarse" channels and then further separates the coarse channels into 128$\times$10\,kHz ``fine" spectral channels.  Observations were taken in two minute segments over a total of 180 minutes, a field of view (primary beam) FWHM of $\sim$30$^{\circ}$ at a resolution (synthesized beam) of 3.2\arcmin.  However, only 625\,deg$^{2}$ were imaged and searched, within the most sensitive region of the primary beam.

The data were imaged and calibrated as part of the molecular line survey of the Orion Molecular Cloud complex (Tremblay et al, in prep) using the pipeline described in \citet{tre17}. By using WSClean \citep{off14} we imaged each coarse and fine spectral channel using Briggs weighting ``-1'' to compromise between image resolution and sensitivity.  Images from each coarse channel were used to determine the effects of the ionosphere on the source positions within the field of view, and a single linear correction was based on a comparison with GaLactic and Extragalactic All-sky Murchison Widefield Array (GLEAM) survey point source catalog \citep{hur17} stacked images at 103\,MHz.  After correction, the residual astrometric uncertainty is 1\arcsec~ in right ascension and 5\arcsec~ in declination; both of which are significantly smaller than our beam.

The edges of each coarse channel suffer from aliasing, requiring a number of fine (10\,kHz) channels on each coarse channel edge to be flagged. This resulted in 2400 (78\%) of the 3072 fine spectral channels being imaged. The central fine channel of each coarse channel was flagged, as they contain the DC component of the filterbank. Automated flagging of radio frequency interference (RFI) was performed using AO Flagger \citep{off12} in each (2-minute) snap-shot observation prior to the visibilities being imaged and stacked.  As noted in \citet{tin16} and \citet{tin15}, the application of AO Flagger detects RFI onn single baselines and over short integration periods, identifying and removing RFI corresponding to thouands of Jy, three to four orders of magnitude higher than the levels being probed in this paper and corresponding to known terrestrial transmitter frequencies (generally in the FM band).  

A search of the field for exoplanets, based on the Kepler catalogue \citep{kep13}, returned 17 known planetary systems containing 22 exoplanets. The systems are listed in Table 2 and are shown in relation to the MWA field of view in Figure 1.

The MWA data cube was searched at the locations of each of these exoplanet systems and spectra were extracted from these locations. No significant narrow band signals were detected above a 5$\sigma$ level in any of these spectra. Table 2 lists the RMS from the spectra at each of the exoplanet system locations and the corresponding 1$\sigma$ limits on the inferred isotropic transmitter power required at the distance of the exoplanet system. Figure 2 shows an example spectrum extracted from the MWA data cube, representing the data for exoplanets BD-06 1339b/c.

\begin{figure*}[ht]
\centering
\includegraphics[width=17cm,angle=0]{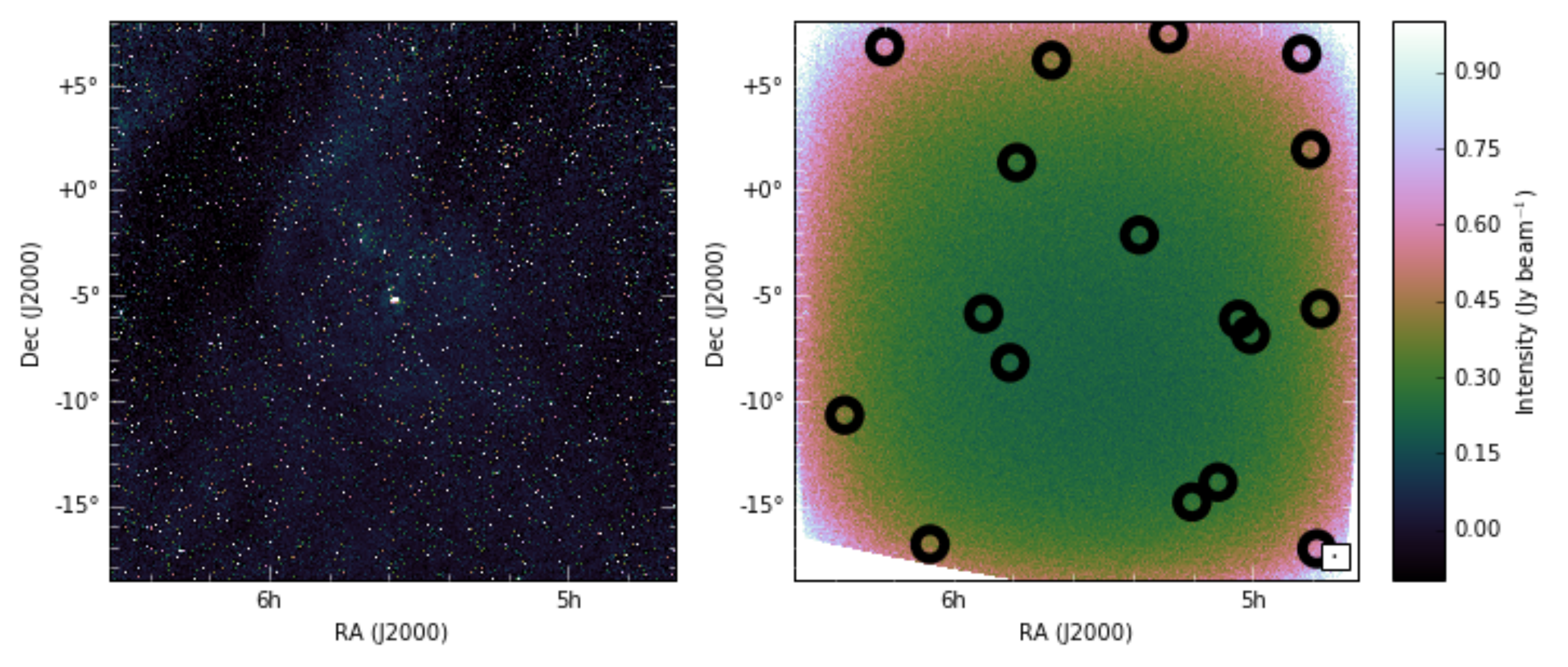}
\caption{Left panel: the continuum image of the Galactic Anticenter field with a central frequency of 114.6 MHz.  Right panel: the distribution of the 22 known exoplanets in the MWA field of view, overlaid on the noise across the field represented by the colorscale, where each pixel represents the RMS along the spectral dimension of the image cubes.}
\end{figure*}

\begin{table}
\small
\caption{MWA Observing Parameters}

\label{obs}
\begin{tabular}{lc}
\hline
Parameter & Value\\
\hline
\hline
Central frequency& 114.56\,MHz\\
Total bandwidth & 30.72\,MHz\\
Number of imaged channels & 2400\\
Channel separation & 10 kHz \\
Synthesized beam FWHM & 3.2$^{\prime}$\\
Primary beam FWHM & 30\, degrees\\
Phase center of map (J2000) &  05h35m, --05d27m\\
Time on source & 3\,hours \\
Date Observed & 21 November 2015\\
						
\hline
\end{tabular}
\end{table}

\begin{table*}[ht]
\centering
\small
  \begin{tabular}{c c c c c c c} \hline 
\# System & RA & Dec & Dist. (pc)&RMS (Jy/beam) & P ($10^{13}$ W) \\ 
BD-06 1339b&05h53m00.28s&$-$05d59m41.4s&	20.1&0.28&$<$14 \\
BD-06 1339c&05h53m00.28s&$-$05d59m41.4s&20.1&0.28&$<$14 \\
GJ 179b&04h52m05.73s&$+$406d28m35.5s&12.12&0.63&$<$11 \\
GJ 3323b&05h01m57.43s&	$-$06d56m46.5s&5.32&0.29&$<$1 \\
GJ 3323c&05h01m57.43s&	$-$06d56m46.5s&5.32&0.29&$<$1 \\
HD 290327b&05h23m21.56s&$-$02d16m39.4s&56.7&0.27&$<$101 \\
HD 30562b&04h48m36.38s&$-$05d40m26.6s&26.5&0.41&$<$34 \\
HD 33142b&05h07m35.54s&$-$13d59m11.3s&136.8&0.31&$<$714 \\		
HD 33844b&05h12m36.10s&$-$14d57m04.0s&100.9&0.33&$<$392 \\
HD 33844c&05h12m36.10s	&$-$14d57m04.0s&100.9&0.33&$<$392 \\
HD 34445b&05h17m40.98s&$+$07d21m12.0s&45.01&0.51&$<$121 \\
HD 37605b&05h40m01.73s&$+$06d03m38.1s&42.88&0.42&$<$91 \\
HD 37605c&05h40m01.73s	&$+$06d03m38.1s&42.88&0.42&$<$91 \\
HD 38529b&05h46m34.91s&$+$01d10m05.5s&42.4&0.31&$<64$ \\		
HD 38529c&05h46m34.91s&$+$01d10m05.5s&42.4&0.31&$<64$ \\
HD 38801b&05h47m59.18s&$-$08d19m39.7s&99.4&0.25&$<$289 \\
HD 42618b&06h12m00.57s&$+$06d46m59.1s&23.5&0.57&$<$37 \\
HD 44219b&06h20m14.32s&$-$10d43m30.0s&50.4&0.43&$<$128 \\
WASP-141	b&04h47m17.86s&$-$17d06m54.6s&570&0.59&$<2.2\times10^{4}$ \\
WASP-35b&05h04m19.63s	&$-$06d13m47.4s&--&0.32&$--$ \\
WASP-49b&06h04m21.46s	&$-$16d57m55.1s&170&0.44&$<1.4\times10^{3}$ \\
WASP-82b&04h50m38.56s	&$+$01d53m38.1s&200&0.46&$<2.1\times10^{3}$ \\ \hline

 \end{tabular}
 \caption{The 22 known exoplanets in the MWA field of view.  Column 1 - Exoplanet system; Column 2 - right ascension (deg); Column 3 - declination (deg);  Column 4 - distance (pc); Column 5 - RMS (Jy); and Column 6 - upper limit on isotropic transmitter power in units of $10^{13}$ W.  The distance of WASP-35b is not known.}
\end{table*}

\begin{figure*}[ht]
\centering
\includegraphics[width=17cm,angle=0]{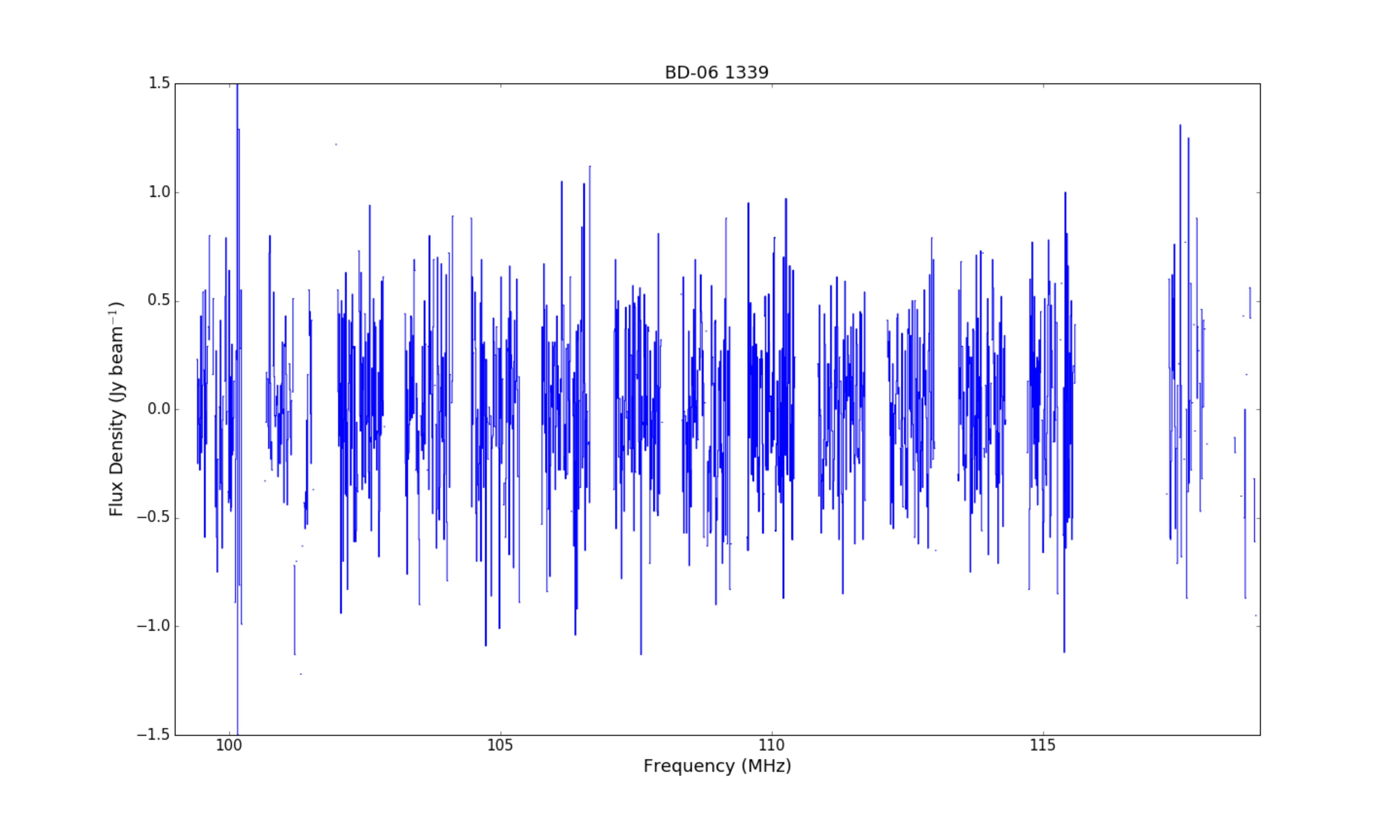}
\caption{MWA spectrum for BD-06 1339b/c.  Areas of flagged fine channels on coarse channel edges are evident as gaps in the spectrum.}
\end{figure*}

\section{DISCUSSION}
GJ 3323b and c, at 5.32 pc, are the closest exoplanets in Table 2, 20\% closer than the closest exoplanet we observed in \citet{tin16}, GJ 667c.  The measured RMS toward GJ 3323b/c is approximately half of the RMS we measured toward GJ 667c.  Thus, our limit on EIRP for GJ 3323b/c is approximately four times lower than we prevsiouly estimated for GJ 667c, $10^{13}$ W.

In terms of the sample in this paper, the median distance ($\sim$50 pc) is a factor of approximately 40 lower than the median distance of the systems in \citet{tin16} ($\sim$2 kpc), meaning a median factor of $\sim$1600 higher EIRP sensitivity in the current sample.

Thus, in terms of the best limits and the overall limits, the observations of the Galactic Anticenter field have yielded more stringent constraints than previous published for our Galactic Center field.

As noted in \citet{tin16}, an EIRP limit of $10^{13}$ W is still high compared to the highest power low fequency transmitters on Earth, even though this EIRP limit is plausibly within reach of Earth-based transmitters at frequencies a factor of ten higher \citep{enr17}.

We note that the primary goals of the observations utilised here are searches for radio recombination lines and molecular lines.  Tremblay et al. 2018a (in preparation) and Tremblay et al. 2018b (in preparation) report both types of astrophysical lines, including three instances of lines that have no identified molecular transition but are plausibly molecular lines rather than SETI signals, given their coincidence with evolved stars.

We note that we only consider explicitly here the positions of known exoplanets.  Across the very large MWA field and the very large numbers of stars this covers, vastly more as yet unknown exoplanets exist.  Our survey is a blind survey across this field, providing non detections toward all exoplanets (known and unknown).  Rather than record a large number of non-detections for thousands of stars here, it is possible for the reader to construct a limit on EIRP based on the position of the stellar system of interest and with reference to the RMS map across the field provided in Figure 1.

\subsection{Ongoing and future work}

The MWA is continuing to build toward larger-scale SETI experiments, in collaboration with the Breakthrough Listen team.

In addition to searches of wide-field image cubes such as the one reported here, the MWA has the capability to beamform using voltages on specific targets of interest. Voltages provided by the Voltage Capture System (VCS; \citealt{mwavcs}) can be processed into data products including incoherent and coherent beams. The former enables a search of a single pixel corresponding to the primary beam of an individual MWA tile; the latter provides (in the case of the MWA) an order of magnitude improvement in sensitivity, for a beam comparable in size to the synthesized beam produced by the correlator. For frequency resolution similar ($\sim 10$\,kHz) to that of the search reported here, forming an image cube using the correlator is tractable in terms of computational expense and data volume, but as frequency resolution improves (and the number of channels increases), beamforming in the direction of targets of interest becomes preferable.

The VCS has been used for fast transient searches and pulsar studies (e.g. \citealt{vcspulsar,vcspulsar2}), among other applications. The two main current limitations to the VCS are the network infrastructure at the Murchison Radio-astronomy Observatory (MRO), limiting the use of the system to just a few hours per week, and the manner in which the data are channelized (limiting the frequency resolution that can be obtained). A new 100 Gbit/s link now operational from MRO to Curtin University overcomes these limitations, and the installation of a new Breakthrough Listen computational facility at Curtin University similar to those deployed at GBT and Parkes \citep{macmahon} will enable a high frequency resolution, real-time commensal SETI search to be performed. This new instrument will not only provide improved SETI search capabilities on MWA, but will enhance the overall capabilities of the telescope for fast transient, pulsar science, molecular line, and solar studies, and is expected to be installed and operational in 2018.

The MWA's wide field of view and broad range of science cases means that a commensal user can build up a deep all-sky survey without ever pointing the telescope. Over the course of a year, almost the entire visible sky is covered to a depth of at least several hours, with some smaller deep fields covered to an order of magnitude or more greater depth. By accessing the data before fine channelization is performed, a substantial improvement over the current 10\,kHz frequency resolution can be obtained, enabling a much more sensitive search for narrow band ($\sim$\,Hz bandwidth) transmitters, as well as improved discrimination between natural spectral lines, RFI, and artificial signals of interest.

\section{Acknowledgements}
We thank the anonymous referee for providing comments that significantly improved the manuscript.  The authors would like to acknowledge the contribution of an Australian Government Research Training Program Scholarship in supporting this research. This work was supported by resources provided by the Pawsey Supercomputing Centre with funding from the Australian Government and the Government of Western Australia.  This scientific work makes use of the Murchison Radio-astronomy Observatory, operated by CSIRO. We acknowledge the Wajarri Yamatji people as the traditional owners of the Observatory site. Support for the operation of the MWA is provided by the Australian Government (NCRIS), under a contract to Curtin University administered by Astronomy Australia Limited.  We gratefully acknowledge the support of NASA and contributors of SkyView surveys.  Funding for Breakthrough Listen research is provided by the Breakthrough Prize Foundation\footnote{\url{http://breakthroughprize.org}}.

{\it Facility:} \facility{MWA}.

\end{document}